\documentclass{ccjnl}
\usepackage{stfloats}
\usepackage{soul, color, xcolor}

\title{Integrated Sensing and Communication for Network-Assisted Full-Duplex Cell-Free Distributed Massive MIMO Systems}
\author{Fan Zeng\inst{1}, Jingxuan Yu\inst{1}, Jiamin Li\inst{1,2,*}, Feiyang Liu\inst{3}, Dongming Wang\inst{1,2}, Xiaohu You\inst{1,2}, \corinfo{jiaminli@seu.edu.cn}}
\receiveddate{Feb.~28,~2023}
\reviseddate{Jul.~09,~2023}
\Editor{}

\address[1]{National Mobile Communications Research Laboratory, School of Information Science and Engineering, Southeast University, Nanjing 210096, China}
\address[2]{Purple Mountain Laboratories, Nanjing 211111, China}
\address[3]{54th Research Institute of Electronics Technology Group Corporation, Shijiazhuang 050081, China}

\begin{document}
	\maketitle
	
	\begin{abstract}
		In this paper, we combine the network-assisted full-duplex (NAFD) technology and distributed radar sensing to implement integrated sensing and communication (ISAC). The ISAC system features both uplink and downlink remote radio units (RRUs) equipped with communication and sensing capabilities. We evaluate the communication and sensing performance of the system using the sum communication rates and the Cramér-Rao lower bound (CRLB), respectively. We compare the performance of the proposed scheme with other ISAC schemes, the result shows that the proposed scheme can provide more stable sensing and better communication performance. Furthermore, we propose two power allocation algorithms to optimize the communication and sensing performance jointly. One algorithm is based on the deep Q-network (DQN) and the other one is based on the non-dominated sorting genetic algorithm II (NSGA-II). The proposed algorithms provide more feasible solutions and achieve better system performance than the equal power allocation algorithm.
	\end{abstract}
	
	\section{introduction}\label{s1}
	Integrated sensing and communication (ISAC)\cite{1,2,3} has emerged as a prominent and challenging research area in the 6G domain due to its multidimensional technical implications and diverse application scenarios, in which communication and radar sensing can effectively improve system spectrum efficiency and hardware resource utilization through time-frequency-space resource reuse and hardware device sharing. In this regard, extensive research efforts have been conducted in the literature to improve sensing and communication performance by proposing innovative designs, such as integrated waveform design\cite{4}, joint transmit beamforming\cite{5}, and joint signal reception\cite{6}. 
	
	While most of the previous works have focused on individual ISAC transceivers, recently, there has been an increasing number of research on networked ISAC receivers. In \cite{7}, a comprehensive overview of sensing mobile networks is provided, proposing the use of multiple ISAC transceivers for distributed radar sensing and coordinated wireless communication. In \cite{8}, a network-integrated sensing and communication system is implemented using multiple ISAC transceivers, where a distributed set of ISAC transmitters send individual messages to their respective communication users while cooperating with multiple sensing receivers to estimate the location of a target. \cite{9} proposes a protocol for communication and sensing, where a group of access points (APs) is allocated by the network to participate in the uplink along with users (UEs). From the perspective of joint communication and radar, a cell-free (CF) massive MIMO network architecture is considered. In \cite{10}, a two-stage sensing framework based on orthogonal frequency-division multiplexing (OFDM) cellular cell architecture is proposed for locating passive targets that cannot send reference signals received from base stations (BSs) to BSs.
	
	Network ISAC offers several advantages over single-cell ISAC, including larger monitoring areas, expanded sensing coverage, diverse sensing angles, and richer sensing information. Furthermore, current research on communication systems is also focused on multiple base stations, so this is in line with the mainstream trend. Additionally, network ISAC can solve the full-duplex problem by transceiver separation than simultaneous transmitting and receiving signal echoes in single-cell ISAC \cite{11,12}. 
	
	However, most of the previous research on network ISAC has primarily focused on integrating sensing with traditional cellular communication systems, such as \cite{8} and \cite{10}. However, both of these assume that the communication signals between each cell are uncorrelated, which is an unreasonable assumption. Additionally, traditional cellular architecture suffers from severe inter-cell interference at the cell edge. A promising technology for future wireless communication standards is the cell-free (CF) communication architecture \cite{13}. By enabling simultaneous coherent multipoint transmission and reception over a large number of spatially distributed remote antenna units (RRUs), this architecture can provide better propagation and channel hardening effects and mitigate cell boundary and inter-cell interference problems associated with cellular systems. Therefore, investigating how to integrate sensing with CF networks is a problem worth exploring.
	
	In the CF network, a significant challenge to reusing communication signals for sensing is the simultaneous transmission of the same data by multiple RRUs to users. The RRU will receive the same signals emitted by many other RRUs, which are all interfering signals. Previous research has shown limited success in extracting useful sensor information from such signals that are superimposed with multiple simultaneous co-channel interferences. Although previous work, such as \cite{9}, has considered it from the CF network perspective, it typically requires a specific access point (AP) for sensing, which limits the overall efficiency of the network. This approach implements sensing and communication separately, and no integration is achieved. 
	
	Furthermore, we observed that the structure of the distributed radar sensing \cite{13} was similar to the network-assisted full-duplex (NAFD) CF distributed massive MIMO network in communication \cite{14,15}. The distributed radar system is a set of radars transmitting signals, and another set of radars receiving echo signals reflected on the target, all of them connected to the CPU. Similarly, NAFD CF distributed massive MIMO network densely distributes multiple RRUs of multiple antennas in a region and connects them to a common CPU performing baseband processing. Each RRU can perform uplink(UL) reception or downlink(DL) transmission in a single time slot, overcoming the problem of full duplex communication interference caused by simultaneous DL transmission and UL reception. To eliminate cross-link interference from DL-RRU to UL-RAU, \cite{16} proposed a beamforming training scheme to help UL-RRU estimate the effective channel state information (CSI).In \cite{17} and \cite{18}, the selection of RRU uplink and downlink modes and power distribution according to user QOS requirements in NAFD scenarios are investigated, respectively. 
	
	Based on the previous analysis, this paper proposes a design method for network ISAC. Referring to the design in \cite{16}, the system uses TDD mode and the first symbol is used for uplink channel estimation, where the CPU estimates the CSI between all users and all RAUs. After completing the uplink channel estimation, the DL-RRUs superimpose the downlink pilot on the downlink data to help the UL-RRUs estimate the effective CSI to eliminate DL-to-UL interference. Simultaneously, UL-RRUs separate the superimposed pilot from the received signal and reuse the superimposed pilot for sensing. The superimposed pilot is used to solve the interference problem caused by multiple RRUs transmitting signals on the same frequency simultaneously in the CF architecture. The superimposed pilot assigned to each DL-RRU is orthogonal, so ideally, there is no interference when sensing. The feasibility of using the superimposed pilot for target sensing has been demonstrated in \cite{19}. The main contributions of this paper are summarized as follows:
	\begin{itemize}
		\item[$\bullet$]Using the design principles of distributed radar sensing, which separates the radar transmitter and receiver, with the structure of separate uplink and downlink remote antenna units (RRUs) in NAFD, we propose a novel design approach for an integrated communication and sensing system with NAFD (NAFD-ISAC). This approach enables a more efficient and effective system by integrating the two structures. To the best of the authors’ knowledge, this kind of system has never been investigated before.
	\end{itemize}
	\begin{itemize}
		\item[$\bullet$]The utilization of the superimposed pilot solves the challenge of interference to sensing caused by multiple RRUs transmitting signals on the same frequency simultaneously in CF architecture. Moreover, it solves the problem of DL-to-UL interference in NAFD, avoiding the impractical assumption of uncorrelated communication signals when reusing communication signals for sensing.
	\end{itemize}
	
	\begin{itemize}
		\item[$\bullet$]Two efficient power allocation schemes, the deep Q-network (DQN) and the non-dominated sorting genetic algorithm II (NSGA-II) are designed from the perspective of multi-objective optimization (MOOP) to achieve better performance of system in communication and sensing.
	\end{itemize}
	\begin{itemize}
		\item[$\bullet$]The conflicting relationship between communication and sensing performance in power allocation is verified, and the results show that there is a trade-off region between communication and sensing. The proposed algorithms provide more feasible solutions and achieve better system performance than the equal power allocation algorithm.
	\end{itemize}
	
	The rest of the paper is organized as follows. Section II models the system from the perspective of a communication system, and radar sensing system, respectively. In Section III, two efficient power allocation schemes are investigated. In Section IV, the simulation results are given and analyzed. The fifth section concludes the paper.
	
	\textbf{Notations}: Bold letters denote vectors or matrices. $\mathbf{I}_M$ denotes an M-dimensional identity matrix. The conjugate transpose and transpose are denoted by $(\cdot)^{\rm H}$ and $(\cdot)^{\rm T}$, respectively. $\left| \cdot \right|$ and $||\cdot ||$ represent the absolute value and spectral norm, respectively. The Kronecker product is denoted by $\otimes$. For the matrix $\mathbf{X}$, the $i$-th row and the $j$-th column of $\mathbf{X}$ are denoted as $\left[ \mathbf{X}\right]_{i,j}$. Matrix inequality $\mathbf{X} \succeq \mathbf{Y}$ denotes that $\mathbf{X} - \mathbf{Y}$ is positive semidefinite. The estimation of $x$ is denoted by $\hat x$, and the estimation error is denoted by $\tilde x$.
	\begin{figure}[!t]
		\centering
		\includegraphics[width=0.5\textwidth]{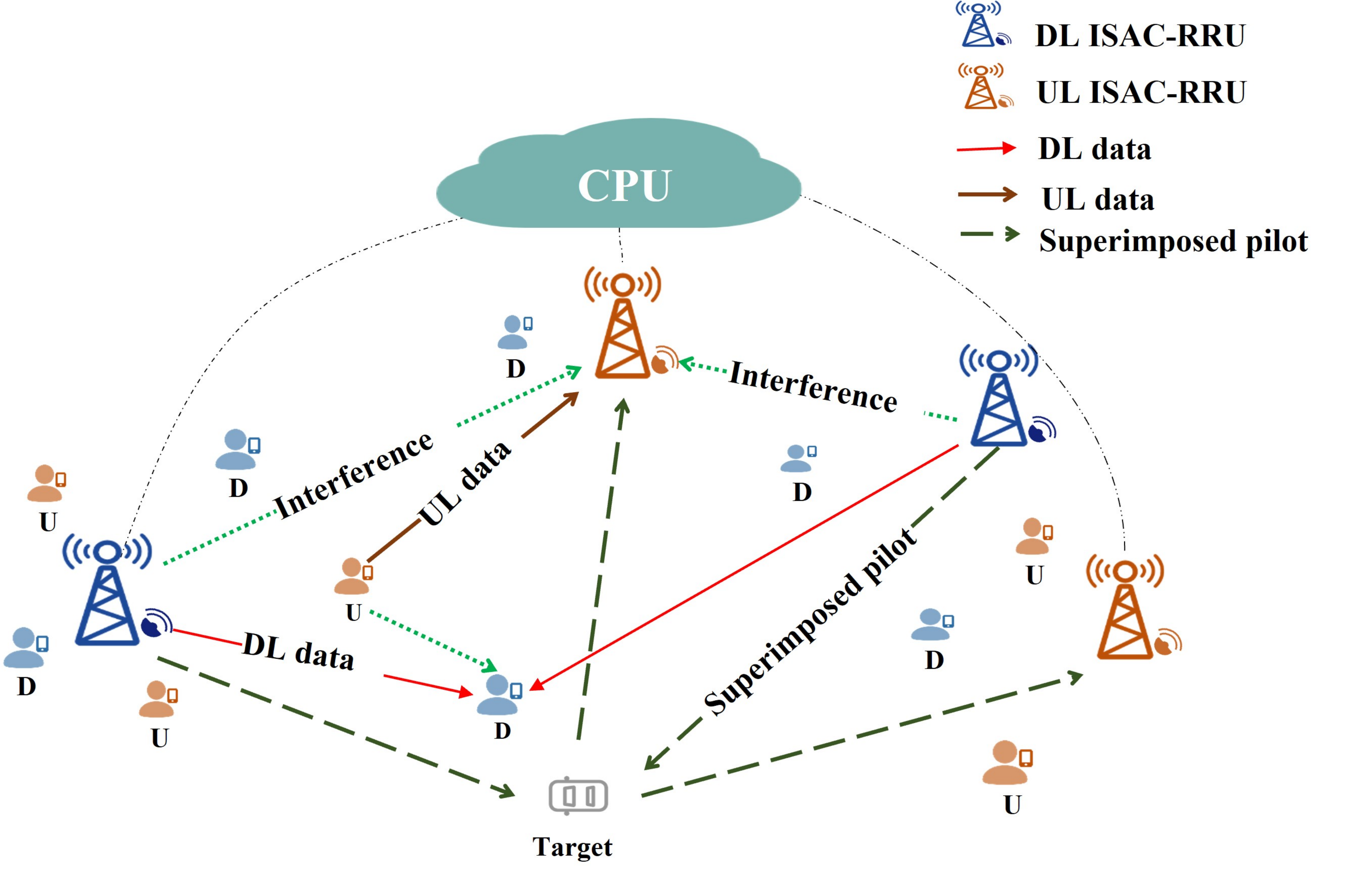}
		\caption{\rm An NAFD-ISAC cell-free distributed Massive MIMO system.}
		\label{fig_1}
	\end{figure}
	\section{System Model}
	\label{s2}
	We consider the NAFD-ISAC CF distributed MIMO system consisting of a CPU, $M$ randomly distributed ISAC-RRU, and $K$ users, a target to be sensed, as shown in Figure~\ref {fig_1}. ISAC-RRUs refer to RRUs with communication and sensing signal processing capability, which we will abbreviate as RRUs later for ease of presentation. ${\kappa _{\rm ul}} = \{ 1,...,{K_{\rm ul}}\}$ and ${\kappa _{\rm dl}} = \{ 1,...,{K_{\rm dl}}\}$ denote the set of uplink users and downlink users, respectively. ${\zeta _{\rm ul}} = \{ 1,...,{M_{\rm ul}}\}$ and ${\zeta _{\rm dl}} = \{ 1,...,{M_{\rm dl}}\}$ denote the set of DL RRUs and UL RRUs, respectively.  The number of antennas of the RRU is $N$, and each user has only a single antenna. The RRU can perform UL reception or DL transmission, which is determined by the CPU. The CPU assigns orthogonal superimposed pilots to the DL-RRUs. We assume that there are ${M_{\rm dl}}$ DL-RRUs, ${M_{\rm ul}}$ UL-RRUs,  ${K_{\rm dl}}$ downlink users and ${K_{\rm ul}}$ uplink users with $M = {M_{\rm dl}} + {M_{\rm ul}}$ and $K = {K_{\rm dl}} + {K_{\rm ul}}$.
	
	First, all users transmit uplink pilot ${\tau _1}$ for uplink channel estimation, and then, in the superimposed pilot and data transmission phase, all DL-RRUs transmit the same data using coherent joint transmission according to a specific beamforming and resource allocation scheme designed by the CPU. At the same time, the CPU assigns to each DL-RRU an orthogonal superimposed pilot and superimposed on the downlink data. The UL-RRUs receive both the uplink data from the UL users and the DL-to-UL interference from the DL-RRUs. The CPU performs channel estimation based on the signals received by the UL-RRUs. After channel estimation, the superimposed pilot is separated from the data and is reused for sensing analysis.
	\subsection{Communication Model}
	From the communication perspective, in downlink data transmission, $M_{\rm dl}$ DL RRUs send data $\boldsymbol{s}$ to $K_{\rm dl}$ downlink users. At each DL RRU, an orthogonal superimposed pilot $\boldsymbol{\varphi}$ is added. The signal transmitted by the $m$-th DL RRU can then be represented as follows:
	\begin{equation}
		\boldsymbol{z}_{\rm dl,m}^{} = \sqrt {{p_{\max }}{\beta _m}}{\mathbf{w}^{s}_{m}}
		\boldsymbol{\varphi} _m^{} + \sum\limits_{i \in {\kappa _{\rm dl}}} {\sqrt {{p_{\max }}{\alpha _{m,i}}} \mathbf{w}^{c}_{i,m}}s_i^{},
	\end{equation}
	where $p_{\max}$ is the maximum signal power that can be transmitted by the DL-RRU, $\alpha_{m, i}$ is the data power ratio factor to the total maximum transmit power transmitted by the DL-RAUs, $\beta_m$ is the pilot power factor to the total maximum transmit power, $\mathbf{w}^{c}_{i,m}$ is the data beamforming vector using ZF beamforming, $\mathbf{w}^{s}_m$ is the sensing beamforming vector using conjugate sensing beamforming\cite{20}, ${\varphi}_m$ is the superimposed pilot assigned to the $m$-th RRU with $\mathbb{E}[{\varphi}_m^{\rm H}{\varphi}_m]=1$, and $s_i$ is the transmitted downlink data signal with $\mathbb{E}[s_i^{\rm H}s_i]=1$. Then, the data received by the $l$-th downlink user can be expressed as
	\begin{align}
		r_l^{\rm dl} = &\sum\limits_{i \in {\kappa _{\rm dl}}} {\underbrace {\sqrt {{p_{\max }}{\alpha _{m,i}}} {\mu^{c}_{l,i}}{s_i}}_{\rm downlink{\text{ }}\rm data} + \sum\limits_{m \in {\zeta _{\rm dl}}} {\underbrace {\sqrt {{p_{\max }}{\beta _m}}{\mu^{s}_{l,m}}{\varphi} _m^{}}_{\rm superimposed{\text{ }}\rm pilot}} } \nonumber\\
		&+\sum\limits_{k \in {\kappa _{\rm ul}}} {\underbrace {\sqrt {{p_{{\rm ul},k}}} {g_{ {\rm t},l,k}}{x_k}}_{\rm uplink{\text{ }}\rm interference}}  + {n_{\rm dl}},
	\end{align}
	where ${\mu^{c}_{l,i}} \buildrel \Delta \over = \mathbf{g}_{{\rm dl},l}^{\rm H}{\mathbf{w}^{c}_i}$ is defined as the effective CSI between the $l$-th DL user and all DL-RRUs, ${\mu^{s} _{l,m}} \buildrel \Delta \over = \mathbf{g}_{{\rm dl},l,m}^{\rm H}{\mathbf{w}^{s}_m}$ is defined as the sensing interference. Here, ${\mathbf{g}_{{\rm dl},l}} = {[\mathbf{g}_{{\rm dl},l,1}^{\rm T},...,\mathbf{g}_{{\rm dl},l,{M_{\rm dl}}}^{\rm T}]^{\rm T}}$ denotes the downlink channel vector between all DL-RRUs and the $l$-th DL user, ${\mathbf{g}_{{\rm dl},l,m}} = \lambda _{{\rm dl},l,m}^{1/2}{\mathbf{h}_{{\rm dl},l,m}}$ denotes the channel between the $l$-th user and the $m$-th DL-RRU, $\lambda _{{\rm dl},l,m}^{1/2} \buildrel \Delta \over = d_{\rm dl,l,m}^{-{\alpha _{\rm dl}}}$ denotes the large scale fading, $d_{{\rm dl},l,m}^{}$ denotes the transmitted distance, ${\alpha _{\rm dl}}$ denotes the path loss, and ${\mathbf{h}_{{\rm dl},l,m}} \sim {\cal C}{\cal N}(0,{I_N}) \in \mathbb{C} {^{N \times 1}} $ is the small scale fading, $g_{{\rm t},l,k}^{} = \lambda _{{\rm dl},l,k}^{1/2}{h_{{\rm dl},l,k}} \in \mathbb{C} {^{1 \times 1}}$ denotes the channel for the $l$-th UL user to the $k$-th DL user, ${x_k}$ is the data sent by the $k$-th uplink user with $\mathbb{E}\left[ {x_k^{\rm H}{x_k}} \right] = 1$, the data sent by the uplink user is an interference to the downlink user, ${n_{\rm dl}} \sim {\cal C}{\cal N}(0,\sigma _{\rm dl}^2)$ is the additive Gaussian white noise with covariance $\sigma _{\rm dl}^2$. 
	
	On the user side, local channel estimation is performed when the signal is received. We can use the iterative method for channel estimation to get the channel between the users and the DL-RRUs. In this paper, for simplicity, we model the channel estimation error as additive Gaussian white noise with covariance $\sigma _{\rm sp,\rm dl}^2$\cite{21,22}. The $l$-th user and the $m$-th DL-RRU which is expressed as $\mathbf{g}_{{\rm dl},l,m}^{} = \mathbf{\hat g}_{{\rm dl},l,m}^{} + \mathbf{\tilde g}_{\rm dl,l,m}^{}$. Here, $\mathbf{\tilde g}_{{\rm dl},l,m}^{}$ denotes the channel estimation error. Once the user channel estimation process is complete, the downlink pilot signal can be recovered by the user, since the pilot is typically known\cite{23,24}. This enables us to separate the pilot and data signals from each other. The rest signal can be expressed as
	\begin{align}
		\bar r_l^{\rm dl} = &\sum\limits_{i \in {\kappa _{\rm dl}}} {\sqrt {{p_{\max }}{\alpha _{m,i}}} {\mu ^{c}_{l,i}{s_i}} + } \sum\limits_{k \in {\kappa _{\rm ul}}} {\sqrt {{p_{\rm ul,k}}} {g_{ {\rm t},l,k}}{x_k}} \nonumber\\
		&+ \sum\limits_{m' \in {\zeta _{\rm dl}}}^{} {\sqrt {{p_{\max }}{\beta _{m'}}}{\tilde \mu^{s} _{l,m}}{\varphi} _{m'}^{}}  + {n_{\rm dl}},
	\end{align}
	then the downlink data transmission rate is expressed as
	\begin{equation}
		{R_{l}^{\rm dl}} = \mathbb{E}\left[ {{{\log }_2}(1 +{{{\rm \gamma} _l^{\rm dl}}})}\right],
		\label{r_dl}
	\end{equation}
	where $\gamma_l^{\rm dl}$ is the signal-to-interference-plus-noise ratio (SINR) of the $l$-th downlink user, which is given by Eq.~\eqref{gamma_dl} on the top of the next page. From Eq.~\eqref{r_dl}, it can be seen that the downlink rate is affected by downlink data power, uplink data power, and channel estimation error interference. All of these are inextricably linked to power allocation. 
	\begin{figure*}
		\begin{align}
			\gamma_l^{\rm dl}=\frac{{\sum\limits_{m \in {\zeta _{\rm dl}}} {{{\left| {\sqrt {{p_{\max }}{\alpha _{m,l}}} {\mu^{c} _{l,l}}} \right|}^2}} }}{{\sum\limits_{i \in {\kappa _{\rm dl}},i \ne l} {\sum\limits_{m \in {\zeta}} {{{\left| {\sqrt {{p_{\max }}{\alpha _{m,i}}} {\mu^{c} _{l,i}}} \right|}^2}} }  + \sum\limits_{k \in {\kappa _{\rm ul}}} {{{\left| {\sqrt {{p_{\rm ul,k}}} {g_{ {\rm t},l,k}}} \right|}^2} +  \sum\limits_{m' \in {\zeta _{\rm dl}}} {{{\left\| {\sqrt {{p_{\max }}{\beta _{m'}}} {\tilde \mu^{s} _{l,m}}} \right\|}^2}}  + \sigma _{\rm dl}^2} }}
			\label{gamma_dl}
		\end{align}
	\end{figure*}
	
	For uplink transmission, all UL-RRUs receive signals from UL users and signals from DL-RRUs, then the received signals can be expressed at the CPU as follows
	\begin{align}
		\boldsymbol{r}_{}^{\rm ul} = &\sum\limits_{i \in {\kappa _{\rm ul}}} {\underbrace {\sqrt {{p_{\rm ul,i}}} {\mathbf{g}_{l,i}}{x_i}}_{\rm uplink{\text{ }}\rm data}}  + \sum\limits_{l \in {\kappa _{dl}}} {\underbrace {\sqrt {{p_{\max }}{\alpha _{m,l}}} {\mathbf{f}_l}{s_l}}_{\rm {DL-to-UL}{\text{ }}\rm interference}}  \nonumber\\
		&+ \sum\limits_{m \in {\zeta _{\rm dl}}} {\underbrace {\sqrt {{p_{\max }}{\beta _m}} {\mathbf{G}_{{\rm I},m}}{\mathbf{w}^{s}_m}{\varphi} _m^{}}_{\rm superimposed{\text{ }}\rm pilot}}  + {\mathbf{n}_{\rm ul}},
		\label{receve uplink data}
	\end{align}
	where $\boldsymbol{r}_{}^{\rm ul} \in\mathbb{C} {^{{M_{\rm ul}}N \times 1}}$, ${\mathbf{f}_l} \buildrel \Delta \over = {\mathbf{G}_{\rm I}}{\mathbf{w}_l}$ is defined as the effective interference between DL-RRUs and UL-RRUs, ${\mathbf{G}_{\rm I}} = {[\mathbf{G}_{{\rm I},1}^{\rm T},...,\mathbf{G}_{{\rm I},{M_{\rm ul}}}^{\rm T}]^{\rm T}}$. ${\mathbf{G}_{{\rm I},l}} = {[\mathbf{G}_{{\rm I},l,1}^{\rm T},...,\mathbf{G}_{{\rm I},l,{M_{\rm ul}}}^{\rm T}]^{\rm T}}$ denotes the $l$-th interfering channel between DL-RRUs and UL-RRUs, where $\mathbf{G}_{{\rm I},l,{M_{\rm ul}}}^{\rm T} = \lambda _{I,l,m}^{1/2}{\mathbf{H}_{{\rm I},l,m}} \in \mathbb{C}{^{N \times N}}$, \({\mathbf{n}_{\rm ul}}\sim {\cal C}{\cal N}(0,\sigma _{\rm ul}^2\mathbf{I}) \in \mathbb{C}{^{{M_{\rm ul}}N \times 1}}\) is the additive Gaussian white noise with covariance $\sigma _{\rm ul}^2$. Similarly, the CPU performs channel estimation based on the received signal to obtain the channel between DL-RRUs and UL-RRUs. $\mathbf{\tilde G}_{{\rm I},m}^{} = \mathbf{G}_{{\rm I},m}^{} - \mathbf{\hat G}_{{\rm I},m}^{}$ denotes the channel estimation error, which is assumed to be additive Gaussian white noise with the covariance of $\sigma _{\rm sp,\rm ul}^2$. CPU can replicate the pilot signal at the receiver using the initial pilot and estimated channel to recover the pilot signal and then separate the pilot signal from data, the remaining signals are expressed as
	\begin{align}
		\boldsymbol{\bar r}_{}^{\rm ul} = &\sum\limits_{i \in {\kappa _{\rm ul}}} {\sqrt {{p_{{\rm ul},i}}} {\mathbf{g}_{l,i}}{x_i}}  + \sum\limits_{l \in {\kappa _{\rm dl}}} {\sqrt {{p_{\max }}{\alpha _{m,l}}} {\mathbf{f}_l}{s_l}}  \nonumber\\
		&+ \sum\limits_{m \in {\zeta _{dl}}} {\sqrt {{p_{\max }}{\beta _m}} {{\mathbf{\tilde G}}_{{\rm I},m}}{\mathbf{w}^{s}_m}{\varphi} _m^{}}  + {\mathbf{n}_{\rm ul}}.
		\label{rest uplink data}
	\end{align}
	
	Furthermore, since the channel $\mathbf{\hat G}_{{\rm I},m}^{}$ and the transmitted signal $\sqrt {{p_{\max }}{\alpha _{m,l}}} {s_l}$ are known, the CPU can also regenerate the DL-to-UL interference and remove it from the received signal. After subtracting regenerative interference, the CPU uses a specific receiver, then the received signal of the $k$-th user is expressed as
	\begin{align}
		\boldsymbol{\bar r}_{k}^{\rm ul} =\boldsymbol{v}_k^{\rm H}\Big(& \sum\limits_{i \in {\kappa _{\rm ul}}} {\sqrt {{p_{{\rm ul},i}}} {\mathbf{g}_{l,i}}{x_i}}  + \sum\limits_{l \in {\kappa _{\rm dl}}} {\sqrt {{p_{\max }}{\alpha _{m,l}}} {\mathbf{\tilde f}_l}{s_l}}  \nonumber\\
		&+ \sum\limits_{m \in {\zeta _{\rm dl}}} {\sqrt {{p_{\max }}{\beta _m}} {{\mathbf{\tilde G}}_{{\rm I},m}}{\mathbf{w}^{s}_m}{\varphi} _m^{}}  + {\mathbf{n}_{\rm ul}}\Big),
	\end{align}
	then the uplink data transmission rate is expressed as 
	\begin{align}
		R_{{\rm ul},k}^{} = \mathbb{E}\left[ {{{\log }_2}(1 + \gamma_k^{\rm ul})}\right],
		\label{r_ul}
	\end{align}
	where $\gamma_k^{\rm ul}$ is the signal-to-interference-plus-noise ratio (SINR) of the $k$-th uplink user, which is given by Eq.~\eqref{gamma_ul} on the top of this page. Similarly, in Eq.~\eqref{r_ul}, the main interference is due to channel estimation error, and the magnitude of the interferences is also affected by power allocation.
	
	\begin{figure*}[ht]\vspace{-0.8cm} 
		\begin{align}
			\gamma_k^{\rm ul}=\frac{{{p_{\rm ul,k}}{{\left| {\mathbf{v}_k^{\rm H}{\mathbf{g}_{l,k}}} \right|}^2}}}{{{{\sum\limits_{i \in {\kappa _{\rm ul}},i \ne k} {{p_{\rm ul,i}}\left| {\mathbf{v}_k^{\rm H}{\mathbf{g}_{l,i}}} \right|} }^2} + {{\sum\limits_{l \in {\kappa _{\rm dl}}} {\left| {\sqrt {{p_{\max }}{\alpha _{m,l}}} \mathbf{v}_k^{\rm H}{{\mathbf{\tilde f}}_l}} \right|} }^2} + \sum\limits_{m \in {\zeta _{\rm dl}}} {{{\left| {\sqrt {{p_{\max }}{\beta _m}} \mathbf{v}_k^{\rm H}{\mathbf{\tilde G}_{{\rm I},m}}}{\mathbf{w}^{s}_m} \right|}^2}}  + \sigma _{\rm ul}^2{{\left\| {\mathbf{v}_k^{\rm H}} \right\|}^2}}}
			\label{gamma_ul}
		\end{align}
	\end{figure*}

	\subsection{Radar Model }
	
	Viewed from the radar sensing side, the DL-RRUs superimpose the pilot on the communication data, and the UL-RRUs separate the superimposed pilot from the received echo signal and perform radar sensing on the target based on the pilot. For clutter interference of sense, we make some idealized assumptions. We assume that the target-free channel between the DL-RRUs and UL-RRUs is acquired before sensing in the absence of the target. Following the previous literature\cite {25,26}, we neglect the paths resulting from multi-reflections from the other objects due to the presence of the target. Since the transmitted signal is also known at the CPU, the target-free part of the received signal at each UL-RRU can be canceled. After this cancellation, the separated pilot signal in the presence of the target at the $n$-th UL-RRU can be expressed as \cite{27}:
	
	\begin{align}
		{\boldsymbol{r}_n} =\sum\limits_{m \in {\zeta_{\rm dl}}}\sqrt {{p_{\max }}{\beta _m}} {\eta _{n,m}}{{\boldsymbol{\nu} _{n,m}}}{\mathbf{w}^{s}_m}{\varphi} _m^{} + {\boldsymbol{w}_n},
	\end{align}
	where ${\eta _{n,m}} =\tfrac{{{\lambda ^2}{G_{\rm t}}{G_{\rm r}}\sigma }}{{{{(4\pi )}^3}d_n^2d_m^2}}{e^{ - j2\pi (\frac{{\Delta f}}{\rm c}{d_{n,m}}+\frac{{{\rm T_p}}}{\lambda }{{\dot d}_{n,m}})}}$ is a complex amplitude, which is assumed to be deterministic and constant throughout the processing interval, ${G_{\rm t}}$, ${G_{\rm r}}$ denotes the transmitting antenna gain, $\sigma$ is the radar cross-sectional area (RCS) of the target, ${d_{n,m}} = {d_n} + {d_m}$ is the bistatic range between the transmitters, scatterer and receivers, ${\dot d_{n,m}} = {\dot d_n} + {\dot d_m}$ is the associated bistatic range rate, ${d_n}$ and ${d_m}$ denote the distance between transmitter and scatterer and scatterer and receiver, respectively, $\Delta f$ is the bandwidth, ${\boldsymbol{\nu} _{n,m}}$ is the spatial steering vector, $N$ is the number of antennas and indicates the relative displacement of each RRU antenna array, and ${\boldsymbol{w}_n}$ is the noise due to channel estimation error, which according to Eq.~\eqref{receve uplink data} and Eq.~\eqref{rest uplink data}, is expressed as ${\boldsymbol{w}_n} = \sum\nolimits_{m \in {\zeta_{\rm dl}}} {\sqrt {{p_{\max }}{\beta _m}} {\boldsymbol{{\tilde G}}_{{\rm I},m,n}}\boldsymbol{\varphi} _m^{}}$. The target of the scenario considered in this paper is slow-moving. It is assumed to be approximately immobile within a time window, so ${\dot d_{n,m}} = 0$.The steering vector ${\boldsymbol{\nu} _{n,m}} = {\boldsymbol{b}_n} \otimes {\boldsymbol{a}_m} \in \mathbb{C} {^{N \times N}}$ represents the relative phase shift of the RRU on the antenna, where ${\boldsymbol{b}_n}$ and ${\boldsymbol{a}_m}$ are the receive and transmit steering vectors, denoted as ${\boldsymbol{b}_n} = [{b_{n,1}},{b_{n,2}}...,{b_{n,N}}]$ and ${\boldsymbol{a}_m} = [{a_{m,1}},{a_{m,2}}...,{a_{m,N}}]$, respectively, and ${b_{n,j}} = {e^{ - j2\pi \frac{{\boldsymbol{k}_n^{\rm T}{\boldsymbol{q}_{nj}}}}{\lambda }}}$, ${a_{m,i}} = {e^{ - j2\pi \frac{{\boldsymbol{k}_m^{\rm T}{\boldsymbol{q}_{mi}}}}{\lambda }}}$. $\boldsymbol{k}_n^{} = {[\cos ({\theta _n}),\sin ({\theta _n})]^{\rm T}}$ and $\boldsymbol{k}_m^{} = {[\cos ({\phi _m}),\sin ({\phi _m})]^{\rm T}}$ represent the wave vectors for the respective transmit and receive paths, where ${\theta _n}$ and ${\phi _m}$ represent the DOA and DOD, respectively. ${\boldsymbol{q}_{nj}} = {[{x_{nj}},{y_{nj}}]^{\rm T}}$  and ${\boldsymbol{q}_{mj}} = {[{x_{mj}},{y_{mj}}]^{\rm T}}$ represent the position of the $j$-th antenna of the $n$-th UL-RRU and the $i$-th antenna of the $m$-th DL-RRU, respectively.
	
	We analyze the sensing performance by the Cramér-Rao lower bound (CRLB) \cite{28}. The squared position error bound (SPEB) and squared orientation error bound (SOEB) are defined for evaluating performance metrics of radar sense. Assuming that the target location is approximately known a priori, the corresponding CRLB can be optimized to improve the accuracy of the real-time estimation \cite{29}. The CRLB is the mean squared error (MSE) matrix of an unbiased estimate of a known deterministic parameter or estimate of a random parameter, which is the inverse of the bayesian fisher information matrix (FIM) as
	\begin{equation}
		\mathbb{E}\left\{ {({\boldsymbol{\hat \gamma }_{n,m}} - {\boldsymbol{\gamma} _{n,m}}){{({\boldsymbol{\hat \gamma }_{n,m}} - {\boldsymbol{\gamma} _{n,m}})}^{\rm T}}} \right\} \succeq \mathbf{J}_{n,m}^{ - 1},
	\end{equation}
	where ${\boldsymbol{\gamma} _{n,m}} = {\left[ {{d_{n,m}},{\theta _n},{\phi _m}} \right]^T}$ is the true value of the parameters and $\mathbf{J}_{n,m}^{} \in \mathbb{C} {^{3 \times 3}}$ is the FIM of the unknown matrix ${\boldsymbol{\gamma} _{n,m}}$. The FIM can be calculated by the following equation
	\begin{equation}
		{\mathbf{J}_{n,m}} = \frac{1}{{\sigma _n^2}}{\sum\limits_{m \in {\zeta_{dl}}} {\left(\frac{{d{\boldsymbol{r}_{n,m}}}}{{d\boldsymbol{\gamma} _{n,m}^{\rm T}}}\right)} ^{\rm H}}\left(\frac{{d{\boldsymbol{r}_{n,m}}}}{{d\boldsymbol{\gamma} _{n,m}^{\rm T}}}\right).
	\end{equation}
	
	The lower bound of the CRLB for each channel parameter is given by the corresponding diagonal term of $\mathbf{J}_{n,m}^{ - 1}$, and the exact derivation can be found in \cite{28}, whose equation is expressed as
	\begin{equation}
		\mathbf{J}_n^{ - 1} = \mathbf{diag}(\sigma _{d,n}^2 , \sigma _{\theta ,n}^2 , \sigma _{\phi ,n}^2)\label{j_n},
	\end{equation}
	where
	\begin{equation}
		\sigma _{d,n}^2 = \frac{1}{{\sum\limits_{m \in {\zeta_{dl}}} \frac{{p_{\max }}{\beta _m}{\pi ^2}{{({\Delta f}/{c})^2}}\eta _{n,m}^2{N^2}}{{\sigma _n^2}}}{{{\left\|\mathbf{w}^{s}_m \right\|}^2}} },
		\label{sigma_dn}
	\end{equation}
	\begin{equation}
		\sigma _{\theta ,n}^2 = \frac{1}{{\sum\limits_{m \in {\zeta_{dl}}} {\frac{{{p_{\max }}{\beta _m}4{\pi ^2}\eta _{n,m}^2N({B_n} - ({1}/{N})A_n^2)}}{{{\lambda ^2}\sigma _n^2}}}{{{\left\| \mathbf{w}^{s}_m \right\|}^2}} }},
		\label{sigma_theta}
	\end{equation}
	\begin{equation}
		\sigma _{\phi ,n}^2 = \frac{1}{{\sum\limits_{m \in {\zeta_{dl}}} {\frac{{{p_{\max }}{\beta _m}4{\pi ^2}\eta _{n,m}^2N({B_m} - ({1}/{N})A_m^2)}}{{{\lambda ^2}\sigma _n^2}}}{{{\left\| \mathbf{w}^{s}_m \right\|}^2}} }},
		\label{sigma_phi}
	\end{equation}
	${A_m} = \sum\nolimits_{i = 1}^N {\left(({y_{mi}}\cos ({\phi _m})- {x_{mi}}\sin ({\phi _m})\right)}$ and ${B_m} = {\sum\nolimits_{i = 1}^N {\left(({y_{mi}}\cos ({\phi _m})- {x_{mi}}\sin ({\phi _m})\right)} ^2}$ are related to the antenna array on the transmitter, ${A_n} = \sum\nolimits_{i = 1}^N {\left({y_{ni}}\cos ({\theta _n})- {x_{ni}}\sin ({\theta _n})\right)}$ and ${B_n} = {\sum\nolimits_{i = 1}^N {\left(({y_{ni}}\cos ({\theta _n})- {x_{ni}}\sin ({\theta _n})\right)} ^2}$ are related to the receiver. The locally estimated ${\boldsymbol{\hat \gamma} _{n,m}}$ is sent to the CPU, where a fusion is carried out to estimate the position of the target. According to Eq.~\eqref{j_n}, we can get the local estimation error at each UL-RRU, the 1st term on the diagonal represents the estimation error of ${d_{n,m}}$, i.e., the position estimation error SPEB. The 2nd and 3rd terms on the diagonal represent the estimation error of the angle ${\theta _n}$, ${\phi _m}$, i.e., the orientation estimation error SOEB. The SPEB and SOEB are respectively defined as
	\begin{equation}
		e_{\rm sp}^{} = \frac{{\sum\nolimits_{n \in {\zeta_{\rm ul}}} {{{\left[ {\mathbf{J}_n^{ - 1}} \right]}_{1,1}}} }}{{{M_{\rm ul}}}}\label{speb},
	\end{equation}
	\begin{equation}
		{e_{\rm so}} = \frac{{\sum\nolimits_{n \in {\zeta_{\rm ul}}} {\sum\nolimits_{i = 2}^3 {{{\left[ {\mathbf{J}_n^{ - 1}} \right]}_{i,i}}} } }}{{{M_{\rm ul}}}}\label{soeb}.
	\end{equation}
	
	Combining the formula Eq.~\eqref{sigma_dn}-\eqref{sigma_phi} and Eq.~\eqref{speb}-\eqref{soeb} it can be found that both SPEB and SOEB are influenced by the pilot power factor ${\beta _m}$.
	\section{Power allocation schemes}
	Based on the analysis presented, it is evident that the power factor plays an important role in both communication and sensing performance. And the requirements for communication performance and sensing performance are different in different scenarios. Therefore, effective allocation of transmit power is critical to optimize the overall system performance. In this section, we propose a multi-objective optimization problem that aims to maximize both communication and sensing performance while satisfying the system constraints. Specifically, our objective is to maximize the communication rate and minimize the localization error. 
	
	The problem of maximizing the sum communication rate can be expressed as
	\begin{equation}
		\mathop {\max }\limits_{{\boldsymbol{\alpha} _{}}}~{f_1} = {\omega _D}\sum\limits_{l = 1}^{{K_{dl}}} {R_{dl,l}^{} + } {\omega _U}\sum\limits_{k = 1}^{{K_{ul}}} {R_{ul,k}^{}}\label{f1},
	\end{equation}
	where $\boldsymbol{\alpha}  = \left[ {{\alpha _1},...,{\alpha _{{M_{dl}}}}} \right]$ denotes the scale factor of data to total power, ${\alpha _m} = \sum\limits_{i \in {\kappa _{dl}}} {{\alpha _{m, i}}} \forall m \in {\zeta_{dl}}$, ${\omega _D}$ and ${\omega _U}$ denote the sum rate weighting factor for downlink and uplink, respectively. 
	
	Since minimizing the positioning error is equivalent to maximizing the inverse of the position error, the second optimization problem can be formulated as
	\begin{equation}
		\mathop {\max }\limits_{{\boldsymbol{\beta} _{}}}~{f_2} = \frac{1}{{{\omega _{sp}}e_{sp}^{} + {\omega _{so}}e_{so}^{}}}\label{f2},
	\end{equation}
	where $\boldsymbol{\beta}  = \left[ {{\beta _1},...,{\beta _{{M_{dl}}}}} \right]$ represents the scaling factor of the pilot to the total power, ${\omega _{sp}}$ and ${\omega _{so}}$ represent the error weighting factors for position and direction, respectively. 
	
	Mathematically, this multi-objective optimization problem can be formulated as
	\begin{equation}
		\mathop {\max }\limits_{\boldsymbol{\alpha} ,\boldsymbol{\beta} } ~\mathbf{f} = {\left[ {{f_1},{f_2}} \right]^T},
		\label{MOEA}
	\end{equation}
	\begin{equation}
		{\rm s.t.}~\sum\limits_{i \in {\kappa _{dl}}} {{\alpha _{m,i}}{{\left\| {\mathbf{w}_{i,m}^c} \right\|}^2} + {\beta _m}} {\left\| {\mathbf{w}_m^s} \right\|^2} \le 1~\forall m \in {\zeta_{dl}}\label{power constraint },
	\end{equation}
	where $\mathbf{f}$ is a vector of the objective functions $f_1$ and $f_2$, and Eq.~\eqref{power constraint } is the power constraint satisfied by the power allocation factor. Eq.~\eqref{MOEA} aims to maximize the system's communication and sensing performance simultaneously. Due to the non-convexity of objects and the presence of multiple optimization variables, traditional optimization techniques face significant challenges in resolving these complexities. Consequently, we propose a DQN-based solution and a NSGA-II based solution to solve the joint optimization problem.
	
		\subsection{Solution based on DQN}
		In the optimization problem presented above, the optimal power allocation is unknown. This makes it a suitable problem for optimization using Deep Q-Network (DQN). DQN is advantageous in discovering how to choose behaviors that maximize rewards and achieve goals. The agents in DQN interact with dynamic environments through repeated observation, action, and reward to find the optimal strategy. To find this optimal strategy, DQN relies on obtaining the action value function of ${Q_\pi }(s, a)$, which can be estimated using a deep neural network (DNN). In this context, we will explore how the application of a DQN-based algorithm can optimize this problem by considering four key components: agent, state, action, and reward.
		
		\begin{itemize}
			\item \textbf{\emph{Agent}}: We use DL-RAUs as the agent for the reinforcement learning framework. Agents make intelligent decisions based on observation of the environment, including the adaptive selection of DL-RAUs downlink data power factor and the pilot power factor. 
			\item \textbf{\emph{State}}: The state $s_t$ of each agent refers to the channel state information (CSI) and the selection of current downlink data power and pilot power coefficient. 
			\item \textbf{\emph{Action}}: The actions of each DQN agent correspond to the optimization variables, which are the data power coefficient and pilot power coefficient in the optimization problem. These coefficients are typically continuous values, while DQN agents usually take discrete actions. Therefore, we discretized the power coefficients so that they could be represented as discrete actions by the DQN agents.
			\item \textbf{\emph{Reward}}: We define the reward as
			\begin{equation}
				{r_t} \buildrel \Delta \over = {f_1}(\boldsymbol{\alpha} ) + b{f_2}(\boldsymbol{\beta} ),
			\end{equation}
			where ${f_1}(\boldsymbol{\alpha})$ in \eqref{f1} denotes the sum communication rate, ${f_2}(\boldsymbol{\beta})$in \eqref{f2} denotes the inverse of the position error, $b$ is a constant that converts two target values to the same order of magnitude.
		\end{itemize}DQN employs a neural network to derive the mapping of state-action Q values. The input to the neural network is the current state, and its output is the Q value corresponding to all available actions in that state. To determine the t-step action, we implement a greedy approach based on the state-action Q value. At each step t, the neural network estimates the value of state-action Q for every possible action in that state. If the random probability is below $\varepsilon $, then we select the action with the highest Q value. Otherwise, we choose an action randomly. To ensure a stable and effective training process, it is important to dynamically adjust the value of $\varepsilon $. In the early stages, setting $\varepsilon $ to a small value allows for more experience accumulation. As training progresses, gradually increasing $\varepsilon$ helps to stabilize the process. After a certain number of iterations, the output of DQN becomes increasingly stable and the optimal power distribution strategy can be determined based on maximizing rewards.
	
	\subsection{Solution based on NSGA-II}
	
	According to the previous system model, when the data power allocation factor is increased, the data power is increased, and consequently, the sum communication rate also increases. However, the pilot power allocation factor is decreased correspondingly to meet the condition satisfied with the data power allocation factor according to Eq.~\eqref{sigma_dn}-\eqref{sigma_phi}, the corresponding positioning error is increased, so the two objectives contradict each other. Based on the concept of Pareto optimality, a multi-objective optimization method is proposed to explore the trade-off between conflicting objectives.  Despite the potential of the DQN power allocation approach to optimize both communication rate and positioning accuracy, there remains a lack of understanding regarding the Pareto boundary that delineates the trade-off between these two objectives.
	\begin{figure}[ htbp ]
		\centering
		\subfloat[\rm circle deployment]{\includegraphics[width=1\linewidth]{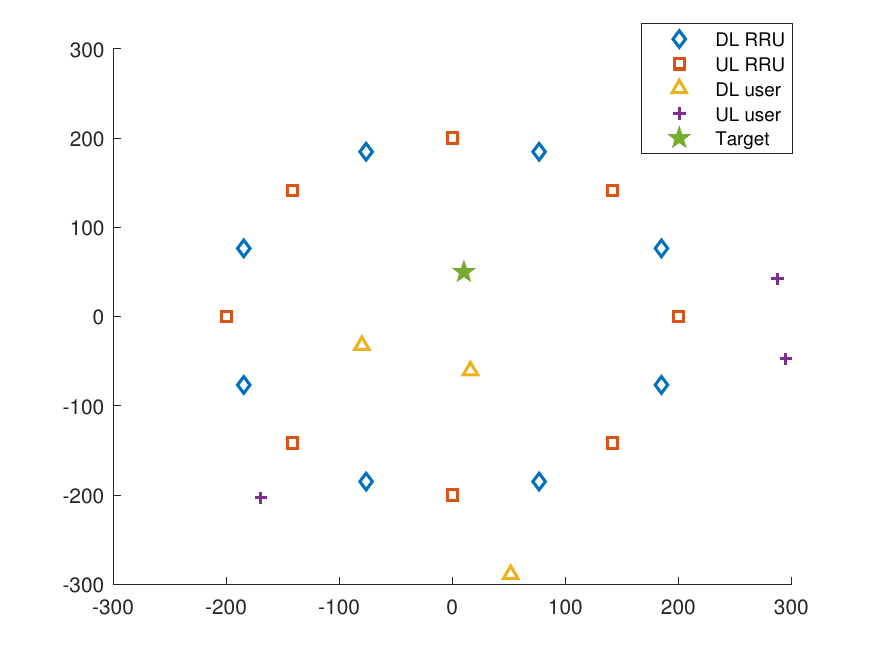}\label{fig2-a}}
		\vspace{-4mm}
		\subfloat[\rm random deployment]{\includegraphics[width=1\linewidth]{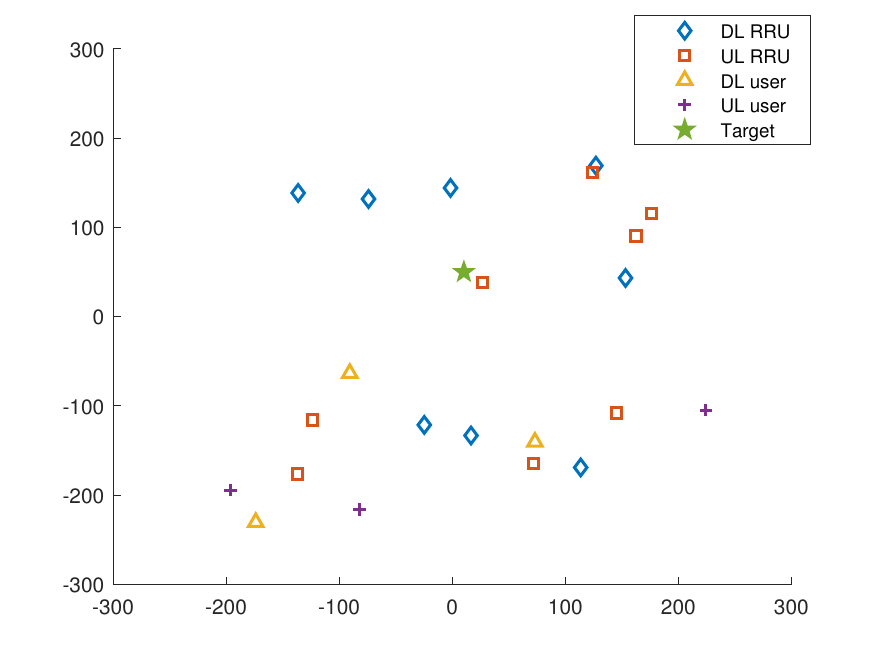}\label{fig2-b}}
		\caption{\rm RRUs, users, and target layout.}
		\label{fig_2}
	\end{figure}
	In this paper, we use multi-objective evolutionary algorithms (MOEAs) to obtain the Pareto bound of Eq.~\eqref{MOEA}. This algorithm has the advantage of solving all objectives simultaneously in one simulation run. Therefore, the MOEAs method is applied in this paper to obtain the Pareto optimal points. Specifically, NSGA-II \cite{30} is used to solve Eq.~\eqref{MOEA}. NSGA-II introduces the crowding degree and the crowding degree comparison operator based on the fast non-dominated sorting method and uses the crowding degree as the comparison criterion among individuals in the population. The individuals of the population in the quasi-pareto region can even be extended to the whole Pareto region, and the good individuals in the population are preserved to ensure the diversity of the population, which has achieved excellent results.
	
	\section{NUMERICAL RESULTS}
	\label{NUMERICAL RESULTS}
	\begin{figure}[ htbp ]
		\centering
		\subfloat[\rm SPEB (m)]{\includegraphics[width=1\linewidth]{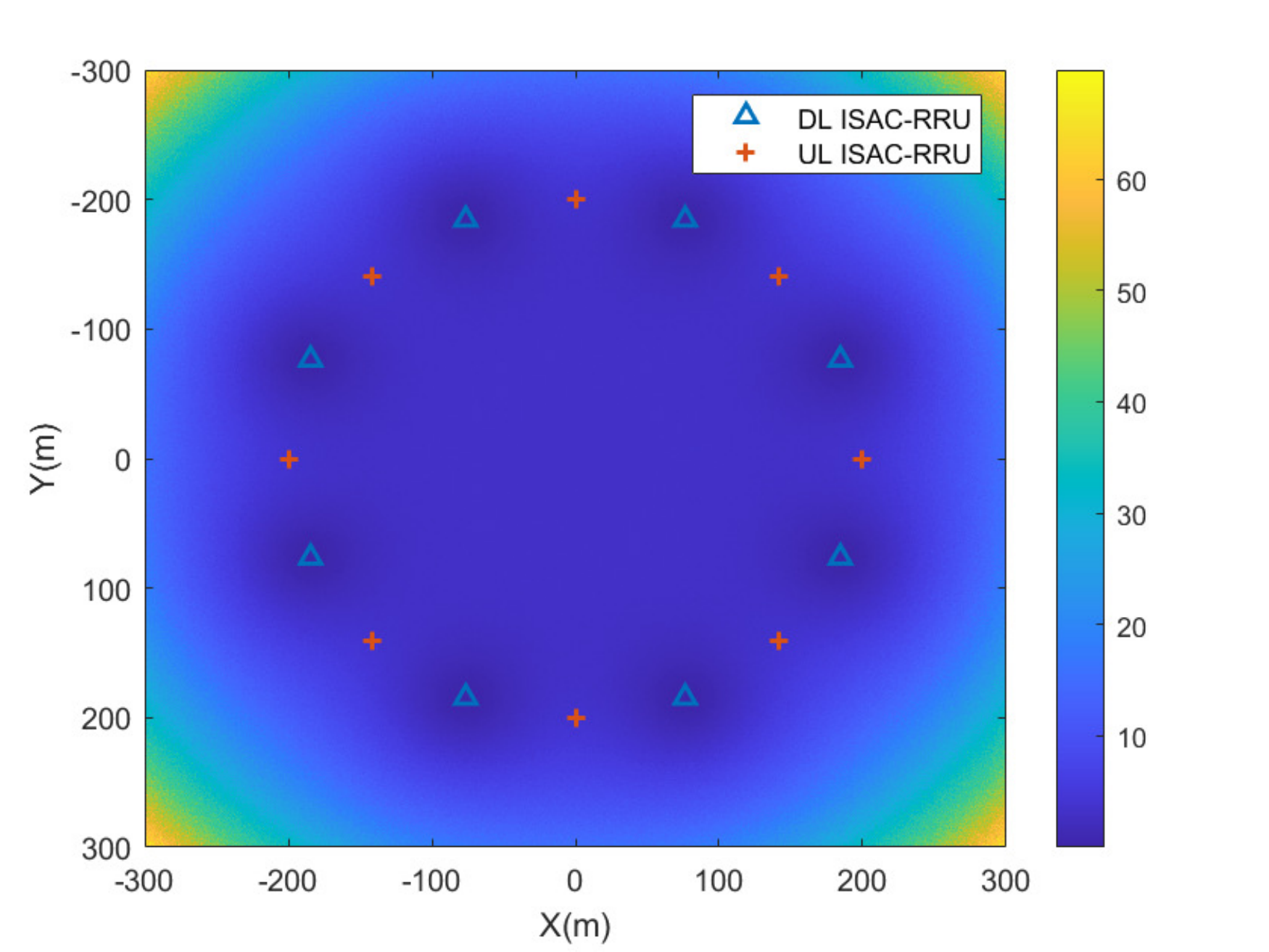}\label{fig_1_a}}\\
		\vspace{-4mm}
		\subfloat[\rm SOEB(deg)]{\includegraphics[width=1\linewidth]{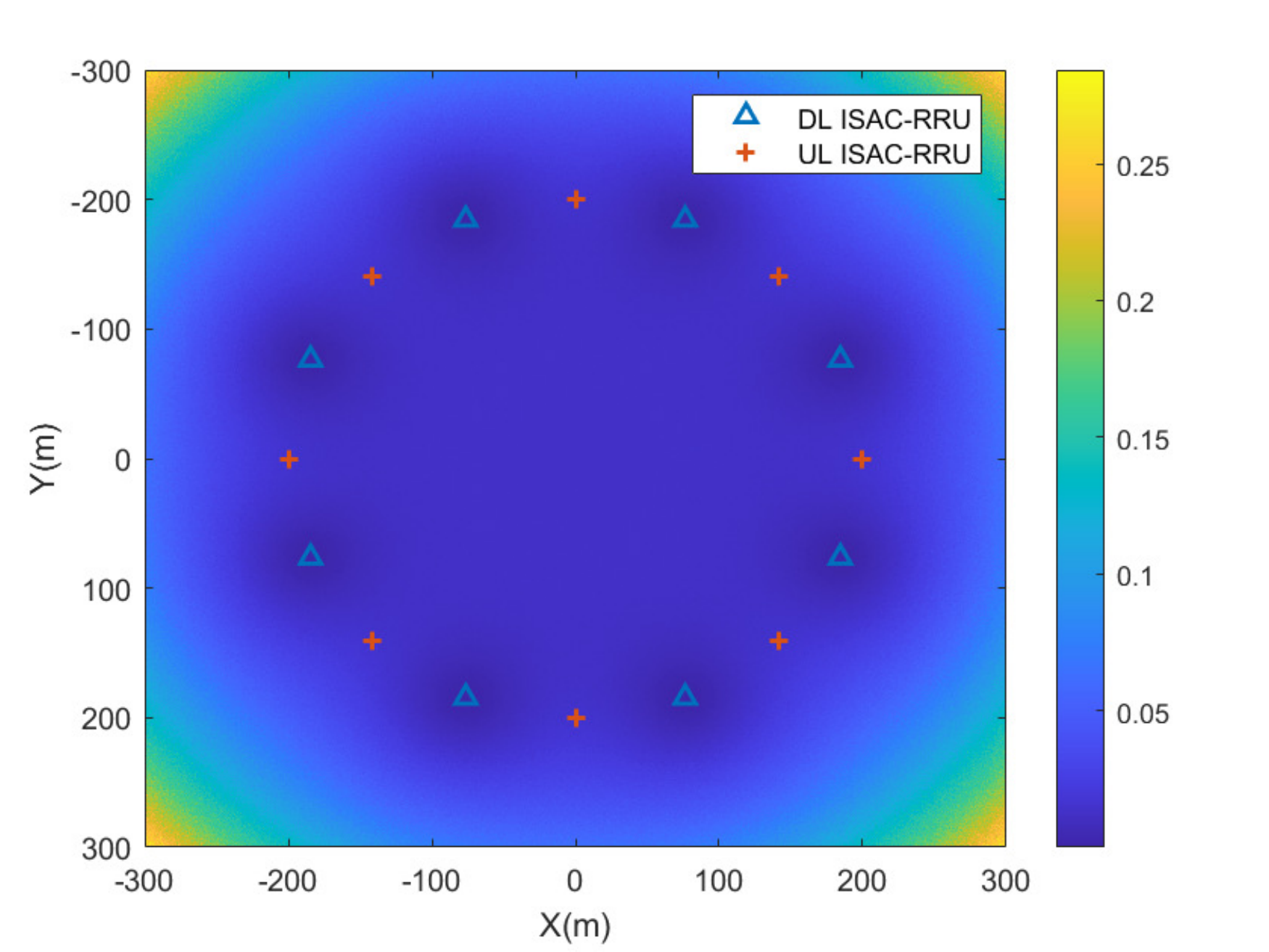}\label{fig_1_b}}\\
		\caption{\rm Contour plots for different metrics in the circle deployment network.}
		\label{fig_3}
	\end{figure}
	In this section, we simulate and analyze the distribution of different sensing performances of the target at different locations and the effect of the power allocation factor on communication and sensing. In addition, the performance of the proposed multi-objective optimization-based power allocation scheme is evaluated.

	Theoretically, the RRU locations are randomly distributed. In the simulation, for the intuition of the results, we assume that there are ${M = 16}$ RRUs uniformly distributed on a circle with a radius of 200 m with $K_{\rm ul}=3$ uplink users, $K_{\rm dl}=3$ downlink users, and a target, randomly distributed in a region of ${R=300}$ m, as shown in Figure~\ref {fig2-a}. To make a partial comparison, we also incorporated the random generation of users, RRU, and target locations within the ${R=300}$ radius area, as depicted in Figure~\ref {fig2-b}. It is assumed that in the NAFD-ISAC distributed MIMO system, half of the RRUs are used for uplink transmission, and the other half for downlink transmission. We set the carrier frequency to 3.5GHz, the bandwidth to 1MHz, and the maximum transmits power of each RRU to 1W. According to \cite{16}, the path loss index is set to ${\alpha _{\rm dl}} = {\alpha _{\rm ul}} = 3.7$, ${\alpha _{\rm t}} = 4$, ${\alpha _{\rm I}} = 3$. The antenna gain is ${G_{\rm t}} = {G_{\rm t}} = 1$ and the RCS is $\sigma  = 1$ . For the noise of the communication signal processing, we choose $\sigma _{\rm dl}^2 = \sigma _{\rm ul}^2 =  - 83$ dBm. The superimposed pilot estimation error is chosen as $\sigma _{\rm sp,dl}^2 = \sigma _{\rm sp,ul}^2 =  - 105$ dBm\cite{31}.
	
	Figure~\ref{fig_3} shows contour plots of various performance metrics for individual targets at different locations in the circle deployment network. To observe the relationship between the location of the target and the estimated error performance, the pilot power factor and the data power factor are set to a and b, and the number of antennas of the RRU is $N=16$. From Figure~\ref{fig_3}, we can see that the performance behaviors of these two metrics have a similar trend with the different locations of the target, and the localization errors are smaller near each DL-RRU, which is related to the RRU layout. When the target is close to a DL-RRU, the signal transmitted by this DL-RAU can be received by two UL-RAUs at approximately the same distance after being reflected by the target, according to the radar complex amplitude equation. The amplitude intensity of the reflected signal increases as the distance between the target and the transmitting and receiving RAUs decreases. Therefore, similar to the previous analysis, there are two RAUs will receive a strong amplitude of the reflected signal. And when the target is near a UL-RAU, the reflected signal will have more losses when received by other UL-RAUs. In addition, we found that the position error is much larger than the orientation error, which is due to the larger number of RAU antennas we placed with higher angular resolution.
	
\begin{figure}[!t]
		\centering
		\includegraphics[width=1\linewidth]{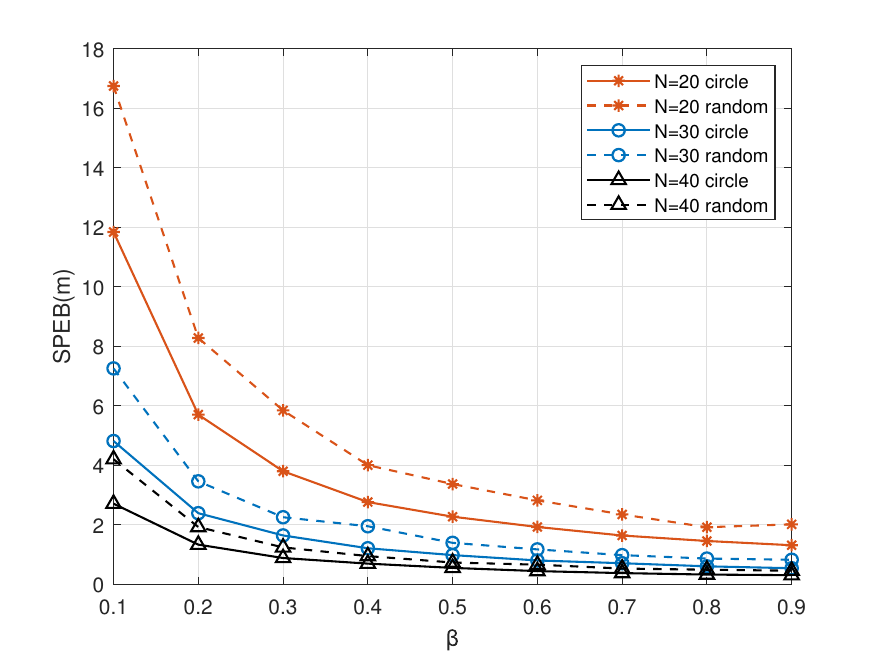}
		\caption{\rm Relationship between SPEB and $\beta_m$ in different deployment.}
		\label{fig_4}
	\end{figure}

	Figure~\ref{fig_4} analyzes the relationship between SPEB and pilot power factor for different numbers of antennas under different network layouts, according to Eq.~\eqref{speb}\eqref{soeb}, SOEB is similar to the relationship between pilot power factor, so only SPEB is simulated here. It can be seen that positioning error is reduced with a higher power factor, and positioning error is reduced with a higher number of aerials. This is the gain of multiple antennas provided by the MIMO system. It can also be observed that as the number of antennas increases, the SPEB of the system becomes less sensitive to varying power for the same network deployment. When comparing the sensing performance of random deployment and circle deployment in a network with equal numbers of antennas, it is found that random deployment exhibits slightly worse results. This result is due to the closer distance between the RRU location and the target in the randomly deployed network in Figure~\ref {fig2-a}. It should be noted that the effectiveness of SPEB may vary depending on the specific network deployment utilized.
	
\begin{figure}[!t]
		\centering
		\includegraphics[width=1\linewidth]{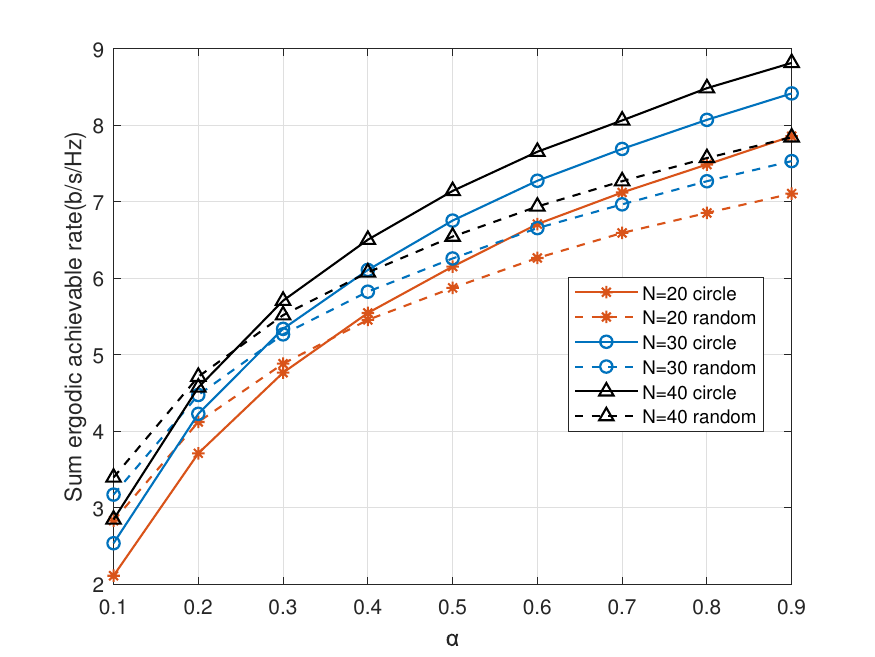}
		\caption{\rm Relationship between sum communication rate and ${\alpha _m}$ in different deployment.}
		\label{fig_5}
	\end{figure}

	Figure~\ref{fig_5} examines the relationship between SPEB and data power factor ${\alpha _m}$ at different numbers of antennas under different network layouts. Under the assumption that the DL-RRU allocates equal power to each user, it can be seen that the data power factor increases with the increase of data power and decreases with the increase of the power factor. This is also a reflection of the need for reasonable power allocation in NAFD-ISAC systems, where a certain amount of power can be allocated for sensing when communication performance is not as high-demand.  Similarly, we can see that the communication performance is better under the random network layout when the power factor of data is relatively small. But with the increase of downlink power, the communication performance decreases compared with circle deployment. This observation can be attributed to the similarity in the distribution of DL-RRUs and DL-UEs positions, along with the proximity of UL-RRUs and UL-UEs positions, leading to a reduction in propagation loss and an improvement in communication performance. As the downlink data power factor increases, the uplink interference also increases gradually, and the uplink communication rate decreases. Therefore, the communication and rate decline compared with circle deployment. Hence, the layout of RRUs plays a crucial role in determining the communication performance of the NAFD-ISAC system. Further research can focus on optimizing RRU placement for maximizing system performance.
	
\begin{figure}[ htbp ]
		\centering
		\subfloat[\rm Sensing performance ]{\includegraphics[width=1\linewidth]{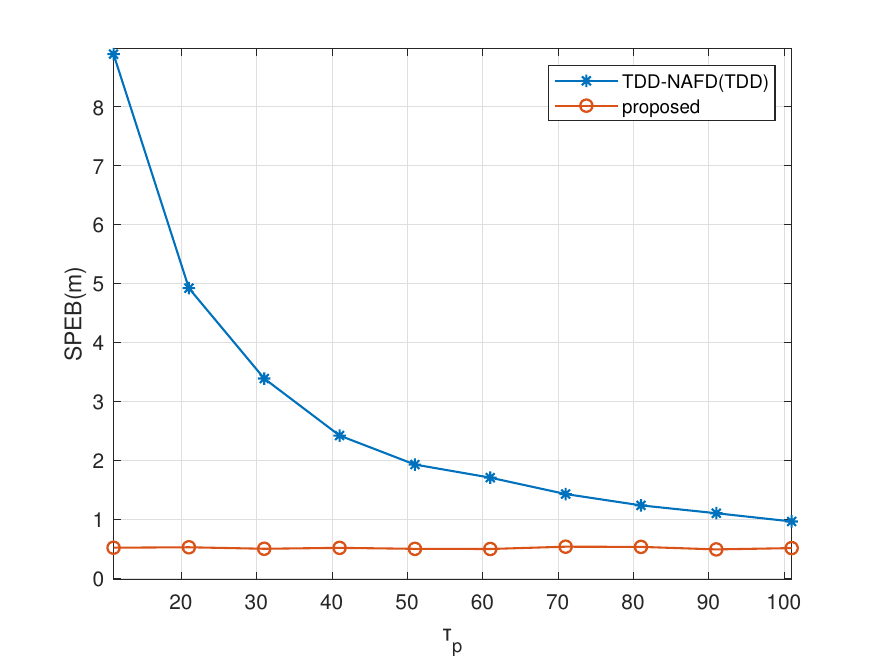}\label{fig_1_a}}\\
		\vspace{-4mm}
		\subfloat[\rm Communication performance]{\includegraphics[width=1\linewidth]{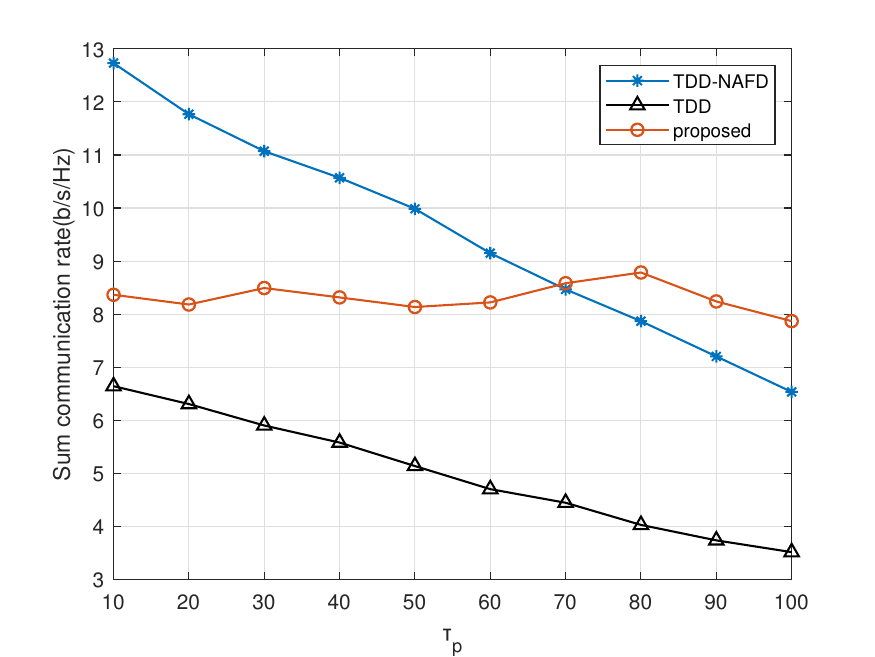}\label{fig_1_b}}\\
		\caption{\rm The performance of different schemes.}
		\label{fig_7}
	\end{figure}

	In Figure ~\ref{fig_7}, we compare the performance of the proposed ISAC system with two baseline schemes with the relationship of sensing duration.
		\begin{itemize}
			\item TDD-ISAC scheme: In this scheme, the sensing service, uplink, and downlink are carried out in turn according to the time slot.
			\item TDD-NAFD-ISAC scheme: This scheme begins by performing the sensing service, followed by employing the NAFD to separate the uplink and downlink RRUs, thereby achieving simultaneous uplink and downlink operations.
		\end{itemize}
		In the scheme proposed in this paper, sensing and communication are carried out at the same time, and the communication within the system adopts NAFD mode. For the proposed scheme, there is no change to the perception duration and it remains a continuous channel block. When comparing perception performance, TDD and TDD-NAFD perform similarly as they are both perceived in a single period. However, the communication performance of TDD-NAFD is superior to that of the TDD and the proposed scheme. TDD-NAFD enables simultaneous uplink and downlink communication which results in improved communication performance compared to TDD. In the proposed scheme, some of the power in the data signal is reserved for sensing, which leads to a slight loss in communication performance but results in better sensing performance. As more sensing symbols are added, the sensing performance will improve, surpassing that of the proposed NAFD scheme. 
		
		However, this also means that the time dedicated to sensing alone will increase, causing a decrease in communication time within a coherent block and ultimately resulting in a decline in communication performance. Based on the previous analysis, we can see that compared with other schemes, the proposed NAFD-ISAC scheme can achieve longer sensing time with stable sensing performance. Although there is a certain loss in communication performance, part of the loss is due to power distribution. The power can be reasonably allocated according to the specific needs of communication performance and sensing performance to achieve better system performance. 
	\begin{figure}[!t]
		\centering
		\includegraphics[width=1\linewidth]{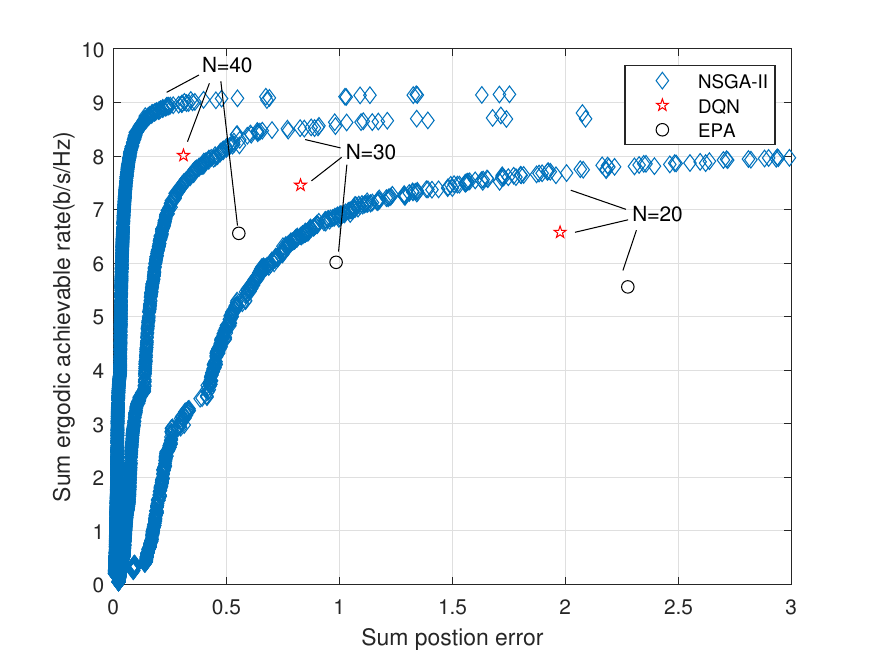}
		\caption{\rm Comparison between equal power allocation(EPA) scheme and proposed scheme.}
		\label{fig_6}
	\end{figure}
	
	The trade-off between the sum communication rate versus location error for different numbers of antennas deployed on each RAU is shown in Figure~\ref{fig_6}. It can be seen that the Pareto front curve is in a monotonically increasing state, and the system positioning error increases as the communication performance increases, while our goal is to minimize the system positioning error, which verifies the contradiction between these two goals in the NAFD-ISAC system. In addition, as the number of antennas increases, the effect on the positioning error becomes smaller and smaller. The solution of the DQN algorithm is better than the equal power allocation(EPA) scheme, the solution obtained by DQN is approximately close to the Pareto solution, and the running complexity of the DQN algorithm is lower than that of the NSGA-II algorithm, but NSGA-II algorithm reveals the optimal frontier of Pareto.

	\section{Conclusion}
	\label{conclusion}
	In this paper, we proposed a method for the design of a network ISAC system, combining the distributed radar with a similar structure NAFD system to design a NAFD-ISAC system. This approach eliminated the full-duplex interference problem in sensing and communication by separating the transceivers. By assigning a superimposed pilot to each DL-RAU, the co-channel interference problem in the CF network is avoided. We simulated the factors influencing the system performance from the communication and sensing perspectives. Furthermore, we proposed two efficient power allocation schemes. The DQN power allocation algorithm solves the problem faster, while the NSGA-II power allocation algorithm provides all the Pareto-optimal solutions. The simulation results proved the proposed algorithms are superior to the equal power allocation. In practical applications, sensing and communication requirements are different. Therefore, how to reasonably select an appropriate RRU distribution and power allocation mode based on communication and sensing requirements in NAFD-ISAC systems is an interesting and challenging problem that deserves attention in future studies.
	
	\section*{ACKNOWLEDGEMENT}
	\label{ACKNOWLEDGEMENT}
	
	This work was supported in part by the National Key Research and Development Program under Grant (2021YFB2900300), and by the National Natural Science Foundation of China (NSFC) under Grants 61971127, 61871122, by the Southeast University-China Mobile Research Institute Joint Innovation Center, and by the Major Key Project of PCL (PCL2021A01-2).
	

	\bibliographystyle{gbt7714-numerical}
	\bibliography{myref}

\begin{thebibliography}{31}
\providecommand{\natexlab}[1]{#1}
\providecommand{\url}[1]{#1}
\expandafter\ifx\csname urlstyle\endcsname\relax\else
  \urlstyle{same}\fi
\expandafter\ifx\csname href\endcsname\relax
  \DeclareUrlCommand\doi{\urlstyle{rm}}
  \def\eprint#1#2{#2}
\else
  \def\doi#1{\href{https://doi.org/#1}{\nolinkurl{#1}}}
  \let\eprint\href
\fi

\bibitem[Cui et~al.(2021)Cui, Liu, Jing, and Mu]{1}
CUI Y, LIU F, JING X, et~al.
\newblock {Integrating} sensing and communications for ubiquitous {IoT}:
  {Applications}, trends, and challenges\allowbreak[J].
\newblock IEEE Network, 2021, 35\allowbreak (5): 158-167.

\bibitem[Wei et~al.(2022)Wei, Liu, Masouros, Su, and Petropulu]{2}
WEI Z, LIU F, MASOUROS C, et~al.
\newblock {Toward} multi-functional {6G} wireless networks: {Integrating}
  sensing, communication, and security\allowbreak[J].
\newblock IEEE Communications Magazine, 2022, 60\allowbreak (4): 65-71.

\bibitem[Xiao et~al.(2022)Xiao and Zeng]{3}
XIAO Z, ZENG Y.
\newblock {An} overview on integrated localization and communication towards
  6g\allowbreak[J].
\newblock Science China Information Sciences, 2022, 65: 1-46.

\bibitem[Liu et~al.(2018)Liu, Zhou, Masouros, Li, Luo, and Petropulu]{4}
LIU F, ZHOU L, MASOUROS C, et~al.
\newblock {Toward} dual-functional radar-communication systems: {Optimal}
  waveform design\allowbreak[J].
\newblock IEEE Transactions on Signal Processing, 2018, 66\allowbreak (16):
  4264-4279.

\bibitem[Sun et~al.(2022)Sun, Dai, and Wang]{5}
SUN P, DAI H, WANG B.
\newblock {Optimal} transmit beamforming for near-field integrated sensing and
  wireless power transfer systems\allowbreak[J].
\newblock Intelligent and Converged Networks, 2022, 3\allowbreak (4): 378-386.

\bibitem[Guo et~al.(2023)Guo, Li, Mei, Yang, Shi, Wong, and Zhang]{6}
GUO T, LI X, MEI M, et~al.
\newblock {Joint} {Communication} and {Sensing} {Design} in {Coal} {Mine}
  {Safety} {Monitoring}: {3D} {Phase} {Beamforming} for {RIS-Assisted}
  {Wireless} {Networks}\allowbreak[J].
\newblock IEEE Internet of Things Journal, 2023.

\bibitem[Zhang et~al.(2020)Zhang, Rahman, Huang, Guo, Chen, and Heath]{7}
ZHANG A, RAHMAN M~L, HUANG X, et~al.
\newblock {Perceptive} mobile networks: {Cellular} networks with radio vision
  via joint communication and radar sensing\allowbreak[J].
\newblock IEEE Vehicular Technology Magazine, 2020, 16\allowbreak (2): 20-30.

\bibitem[Huang et~al.(2022)Huang, Fang, Li, and Xu]{8}
HUANG Y, FANG Y, LI X, et~al.
\newblock {Coordinated} power control for network integrated sensing and
  communication\allowbreak[J].
\newblock IEEE Transactions on Vehicular Technology, 2022, 71\allowbreak (12):
  13361-13365.

\bibitem[Sakhnini et~al.(2022)Sakhnini, Guenach, Bourdoux, Sahli, and
  Pollin]{9}
SAKHNINI A, GUENACH M, BOURDOUX A, et~al.
\newblock {A Target Detection Analysis} in {Cell-Free Massive MIMO Joint
  Communication and Radar Systems}\allowbreak[C]//\allowbreak
ICC 2022-IEEE International Conference on Communications.
\newblock IEEE, 2022: 2567-2572.

\bibitem[Shi et~al.(2022)Shi, Liu, Zhang, and Cui]{10}
SHI Q, LIU L, ZHANG S, et~al.
\newblock {Device}-free sensing in {OFDM} cellular network\allowbreak[J].
\newblock IEEE Journal on Selected Areas in Communications, 2022, 40\allowbreak
  (6): 1838-1853.

\bibitem[Xiao et~al.(2021)Xiao and Zeng]{11}
XIAO Z, ZENG Y.
\newblock {Full}-duplex integrated sensing and communication: {Waveform} design
  and performance analysis\allowbreak[C]//\allowbreak
2021 13th International Conference on Wireless Communications and Signal
  Processing (WCSP).
\newblock IEEE, 2021: 1-5.

\bibitem[Xiao et~al.(2022)Xiao and Zeng]{12}
XIAO Z, ZENG Y.
\newblock {Waveform} design and performance analysis for full-duplex integrated
  sensing and communication\allowbreak[J].
\newblock IEEE Journal on Selected Areas in Communications, 2022, 40\allowbreak
  (6): 1823-1837.

\bibitem[Demir et~al.(2021)Demir, Bj{\"o}rnson, Sanguinetti, et~al.]{13}
DEMIR {\"O}~T, BJ{\"O}RNSON E, SANGUINETTI L, et~al.
\newblock {Foundations} of user-centric cell-free massive {MIMO}\allowbreak[J].
\newblock Foundations and Trends{\textregistered} in Signal Processing, 2021,
  14\allowbreak (3-4): 162-472.

\bibitem[Liang et~al.(2011)Liang and Liang]{14}
LIANG J, LIANG Q.
\newblock {Design} and analysis of distributed radar sensor
  networks\allowbreak[J].
\newblock IEEE Transactions on Parallel and Distributed Systems, 2011,
  22\allowbreak (11): 1926-1933.

\bibitem[Wang et~al.(2019)Wang, Wang, Zhu, Li, Wang, and You]{15}
WANG D, WANG M, ZHU P, et~al.
\newblock {Performance} of network-assisted full-duplex for cell-free massive
  {MIMO}\allowbreak[J].
\newblock IEEE Transactions on Communications, 2019, 68\allowbreak (3):
  1464-1478.

\bibitem[Li et~al.(2020)Li, Lv, Zhu, Wang, Wang, and You]{16}
LI J, LV Q, ZHU P, et~al.
\newblock {Network}-assisted full-duplex distributed massive {MIMO} systems
  with beamforming training based {CSI} estimation\allowbreak[J].
\newblock IEEE Transactions on Wireless Communications, 2020, 20\allowbreak
  (4): 2190-2204.

\bibitem[Zhu et~al.(2022)Zhu, Li, Zhu, Wang, Ye, and You]{17}
ZHU Y, LI J, ZHU P, et~al.
\newblock {Load}-aware dynamic mode selection for network-assisted full-duplex
  cell-free large-scale distributed {MIMO} systems\allowbreak[J].
\newblock IEEE Access, 2022, 10: 22301-22310.

\bibitem[Fan et~al.(2022)Fan, Zhang, Wang, Li, Zhu, and Wang]{18}
FAN Q, ZHANG Y, WANG Z, et~al.
\newblock {MADDPG-Based Power Allocation Algorithm} for {Network-Assisted
  Full-Duplex Cell-Free MmWave Massive MIMO Systems} with {DAC}
  {Quantization}\allowbreak[C]//\allowbreak
2022 14th International Conference on Wireless Communications and Signal
  Processing (WCSP).
\newblock IEEE, 2022: 556-561.

\bibitem[Bao et~al.(2020)Bao, Qin, and Dong]{19}
BAO D, QIN G, DONG Y~Y.
\newblock {A} superimposed pilot-based integrated radar and communication
  system\allowbreak[J].
\newblock IEEE Access, 2020, 8: 11520-11533.

\bibitem[Demirhan et~al.(2023)Demirhan and Alkhateeb]{20}
DEMIRHAN U, ALKHATEEB A.
\newblock {Cell-Free ISAC MIMO Systems: Joint Sensing} and {Communication
  Beamforming}\allowbreak[A].
\newblock 2023.
\newblock arXiv: \eprint{https://arxiv.org/abs/2301.11328}{2301.11328}.

\bibitem[Mishra et~al.(2021)Mishra, Singh, Prasad, and Budhiraja]{21}
MISHRA H~B, SINGH P, PRASAD A~K, et~al.
\newblock {Iterative Channel Estimation And Data Detection} in {OTFS Using
  Superimposed Pilots}\allowbreak[C]//\allowbreak
2021 IEEE International Conference on Communications Workshops (ICC Workshops).
\newblock 2021: 1-6.

\bibitem[Garg et~al.(2022)Garg and Ratnarajah]{22}
GARG N, RATNARAJAH T.
\newblock {Generalized Superimposed Training Scheme In Cell-free Massive MIMO
  Systems}\allowbreak[J].
\newblock IEEE Transactions on Wireless Communications, 2022, 21\allowbreak
  (9): 7668-7681.

\bibitem[Wu et~al.(2023)Wu, Han, and Chen]{23}
WU Y, HAN C, CHEN Z.
\newblock {DFT-Spread Orthogonal Time Frequency Space System} with
  {Superimposed Pilots} for {Terahertz Integrated Sensing} and
  {Communication}\allowbreak[J].
\newblock IEEE Transactions on Wireless Communications, 2023: 1-1.

\bibitem[Deng et~al.(2023)Deng, Fang, and Wang]{24}
DENG C, FANG X, WANG X.
\newblock {Beamforming Design} and {Trajectory Optimization} for {UAV-Empowered
  Adaptable Integrated Sensing} and {Communication}\allowbreak[J].
\newblock IEEE Transactions on Wireless Communications, 2023: 1-1.

\bibitem[Zhao et~al.(2022)Zhao, Wang, Zhang, Chang, and Shen]{25}
ZHAO N, WANG Y, ZHANG Z, et~al.
\newblock {Joint Transmit} and {Receive Beamforming Design} for {Integrated
  Sensing} and {Communication}\allowbreak[J].
\newblock IEEE Communications Letters, 2022, 26\allowbreak (3): 662-666.

\bibitem[Behdad et~al.(2022)Behdad, Demir, Sung, Bj{\"o}rnson, and Cavdar]{26}
BEHDAD Z, DEMIR {\"O}~T, SUNG K~W, et~al.
\newblock {Power Allocation for Joint Communication and Sensing in Cell-Free
  Massive MIMO}\allowbreak[C]//\allowbreak
GLOBECOM 2022 - 2022 IEEE Global Communications Conference.
\newblock 2022: 4081-4086.

\bibitem[Sakhnini et~al.(2021)Sakhnini, Guenach, Bourdoux, and Pollin]{27}
SAKHNINI A, GUENACH M, BOURDOUX A, et~al.
\newblock {A Cram{\'e}r}-rao lower bound for analyzing the localization
  performance of a multistatic joint radar-communication
  system\allowbreak[C]//\allowbreak
2021 1st IEEE International Online Symposium on Joint Communications \& Sensing
  (JC\&S).
\newblock IEEE, 2021: 1-5.

\bibitem[Kwon et~al.(2021)Kwon, Conti, Park, and Win]{28}
KWON G, CONTI A, PARK H, et~al.
\newblock {Joint} communication and localization in millimeter wave
  networks\allowbreak[J].
\newblock IEEE Journal of Selected Topics in Signal Processing, 2021,
  15\allowbreak (6): 1439-1454.

\bibitem[Liu et~al.(2021)Liu, Liu, Li, Masouros, and Eldar]{29}
LIU F, LIU Y~F, LI A, et~al.
\newblock {Cram}{\'e}r-rao bound optimization for joint radar-communication
  beamforming\allowbreak[J].
\newblock IEEE Transactions on Signal Processing, 2021, 70: 240-253.

\bibitem[Qatab et~al.(2018)Qatab, Alias, and Ku]{30}
QATAB W~S, ALIAS M~Y, KU I.
\newblock {Optimization} of multi-objective resource allocation problem in
  cognitive radio {LTE/LTE-A} femtocell networks using {NSGA}
  {II}\allowbreak[C]//\allowbreak
2018 IEEE 4th International Symposium on Telecommunication Technologies (ISTT).
\newblock IEEE, 2018: 1-6.

\bibitem[Hoeher et~al.(1999)Hoeher and Tufvesson]{31}
HOEHER P, TUFVESSON F.
\newblock {Channel} estimation with superimposed pilot
  sequence\allowbreak[C]//\allowbreak
Seamless Interconnection for Universal Services. Global Telecommunications
  Conference. GLOBECOM'99.(Cat. No. 99CH37042): volume~4.
\newblock IEEE, 1999: 2162-2166.

\end{thebibliography}
	
	\biographies
	
	\begin{CCJNLbiography}{authorZeng.eps}{Fan Zeng}
		received the B.S. degree in communication engineering from Shanghai University, Shanghai, China, in 2021. She is currently pursuing the M.S. degree with the College of Information Science and Engineering, Southeast University. Her research interests include integrated sensing and communication. Her supervisor is Professor Li Jiamin.
	\end{CCJNLbiography}
	
	\begin{CCJNLbiography}{authorYu.eps}{Jingxuan Yu}
		received the B.S degree in Information Engineer from Southeast University in 2021 and is currently pursuing his M.S degree in the School of Information Science and Engineering at Southeast University. His main research interests are in communication and sensing integration. His supervisor is Professor Li Jiamin.
	\end{CCJNLbiography}
	
	\begin{CCJNLbiography}{authorLi.eps}{Jiamin Li} received the B.S. and M.S. degrees in communication and information systems from Hohai University, Nanjing, China, in 2006 and 2009, respectively, and the Ph.D. degree in information and communication engineering from Southeast University, Nanjing, China, in 2014. He joined the National Mobile Communications Research Laboratory, Southeast University, in 2014, where he has been an Associate Professor since 2019. His research interests include massive MIMO, distributed antenna systems, and multi-objective optimization.
	\end{CCJNLbiography}
	
	\begin{CCJNLbiography}{authorLiu.eps}{Feiyang Liu}
		received the B.S. and M.S. degrees from Xidian University, Xi'an, China, in 2014 and 2017, respectively. He joined in the 54th Research Institute of China Electronics Technology Group Corporation, in 2017.His research interests include UAV broadband data chain.
	\end{CCJNLbiography}
	\begin{CCJNLbiography}{authorWang.eps}{Dongming Wang}
		received the B.S. degree from Chongqing University of Posts and Telecommunications, Chongqing, China, the M.S. degree from Nanjing University of Posts and Telecommunications, Nanjing, China, and the Ph.D. degree from the Southeast University, Nanjing, China, in 1999, 2002, and 2006, respectively. He joined the National Mobile Communications Research Laboratory, Southeast University, in 2006, where he has been an Associate Professor since 2010. His research interests include turbo detection, channel estimation, distributed antenna systems, and large-scale MIMO systems.
	\end{CCJNLbiography}
	
	\begin{CCJNLbiography}{authorYou.eps}{Xiaohu You}
		received the B.S., M.S. and Ph.D. degrees in electrical engineering from Nanjing Institute of Technology, Nanjing, China, in 1982, 1985, and 1989, respectively. From 1987 to 1989, he was with Nanjing Institute of Technology as a Lecturer. From 1990 to the present time, he has been with Southeast University, first as an Associate Professor and later as a Professor. His research interests include mobile communications, adaptive signal processing, and artificial neural networks with applications to communications and biomedical engineering. He is the Chief of the Technical Group of China 3G/B3G Mobile Communication R \& D Project. He received the excellent paper prize from the China Institute of Communications in 1987 and the Elite Outstanding Young Teacher Awards from Southeast University in 1990, 1991, and 1993. He was also a recipient of the 1989 Young Teacher Award of Fok Ying Tung Education Foundation, State Education Commission of China.
	\end{CCJNLbiography}
\end{document}